# A dosimetric and robustness analysis of Proton Arc Therapy (PAT) with Early Energy Layer and Spot Assignment (ELSA) for lung cancer versus conventional Intensity modulated therapy (IMPT)

Authors: *Macarena S. Chocan [1], Sophie Wuyckens [1], Damien Dasnoy [2], Dario Di Perri [5], Elena Borderias Villarruel [1], Erik Engwall [6], John A. Lee[1], Ana M. Barragán-Montero [1], Edmond Sterpin [1 3 4]

*corresponding author, macarena.chocan@uclouvain.be, https://orcid.org/0000-0001-9697-5440

[1] Université catholique de Louvain, Institut de recherche expérimentale et clinique, Molecular Imaging and Radiation Oncology (MIRO) Laboratory, Brussels, Belgium

[2] Université catholique de Louvain, Institute of Information and Communication Technologies (ICTEAM), Louvain-La-Neuve, Belgium

[3] KULeuven, Department of Oncology, Laboratory of external radiotherapy, Leuven, Belgium

[4] Particle Therapy Interuniversity Center Leuven – PARTICLE, Leuven, Belgium

[5] Department of Radiation Oncology, Cliniques Universitaires Saint-Luc, Brussels, Belgium.

[6] RaySearch Laboratories - Research and Development Department, Stockholm, Sweden

# 1. Introduction

Intensity Modulated Proton Therapy (IMPT) holds promise as a superior choice for radiation treatment in the lung, in comparison with photon-based techniques [1][2][3]. This is primarily attributed to its enhanced dose conformity and absence of exit dose, resulting in improved sparing of organs at risk (OAR). However, IMPT raises significant challenges in treating moving targets [1][4][5], due to the Bragg peak sensitivity to density changes and interplay effect between breathing and the scanned beam.

Proton arc therapy (PAT), a cutting-edge technique that emerges as an alternative to IMPT, involves rotating the proton beam around the patient, either continuously (i.e., dynamic modality)[6] [7] or in a step-and-shoot fashion (i.e., static) [8]. The availability of more degrees of freedom for treatment planning enables potentially superior dose distribution and simplified treatment delivery workflow. However, the total amount of energy layers (ELs) in the plan must be limited to avoid too long and unfeasible treatment plan deliveries. As a result, prior energy layer selection and filtration algorithms have been proposed to address this issue [6] [9] [10] [11] [12], among which the early energy layer and spot assignment algorithm (ELSA), integrated into the commercial treatment planning system (TPS) RayStation (RaySearch Laboratories, Stockholm) [7] ELSA selects a single energy layer (EL) per beam direction from a predetermined set of ELs. Once the optimal EL set is determined, spots are exclusively assigned to those layers, reducing the number of spots to be optimized and thereby decreasing optimization time.

Some planning studies showed a potential benefit for PAT. First, it could preserve target coverage while enhancing dose conformity and OARs sparing [13][14][15], especially for small structures close to the target volume. Second, PAT could achieve a lower body integral dose [13][15][16][17]. Third, PAT could reduce both beam delivery time (BDT) and total treatment time (TTT) due to a simplified treatment workflow. This workflow eliminates the need for couch rotations, manual intervention from technicians to rotate the gantry, and manual beam loading. Additionally, the continuous gantry motion and the absence of lag time between beams contribute to the reduced treatment time. The increased degrees of freedom in PAT could potentially help to improve conformity and better meet clinical objectives. Regarding PAT plan robustness and treatment uncertainties, several studies highlight the importance of a comprehensive robustness analysis for moving tumors [13][14][17][18], in order to evaluate the impact of setup and range uncertainties, acknowledging interplay effects as well.

This study aims to comprehensively analyze potential dosimetric differences and plan robustness in ELSA-based PAT versus conventional IMPT for lung cancer. We use ELSA to create PAT plans and compare them to their IMPT counterparts, in terms of plan quality, normal tissue complication probabilities (NTCP), plan robustness against setup and range uncertainties, and beam delivery time. Additionally, we present results of interplay simulations for both modalities, assessing how intrafraction movement can affect dose coverage on CTV during IMPT and PAT delivery.

# 2. Methods and materials

## 2.1 Patient cohort

A retrospective database of 14 patients with unresectable lung cancer was used. More information can be found in Table SM1. For each patient, a PAT plan and a conventional IMPT plan were created (See section 2.3).

## 2.2 Imaging and target delineation

For each patient, a 4D-CT scan containing 10 phases was acquired. Regular breathing was ensured through audio-coaching [19]. The time-weighted mid position (MidP) image was reconstructed from the 4D-CT scan, using in-house software (REGGUI) [20], and chosen as the planning image[21]. More details about image acquisition, patient immobilization, and MidPCT reconstruction can be found in [19][22].

The gross tumor volume (GTV) was manually delineated by the same radiation oncologist on the MidPCT, while the clinical tumor volume (CTV) was created as an isotropic 5 mm expansion of the GTV [23]. Phases such as end inhale (EndInh), end exhale (EndExh) and mid ventilation (MidV) were extracted from the 4D-CT image series.

## 2.3 Treatment Planning strategy

PAT plans were designed on RayStation research version 12B. Patient-specific IMPT plans had been previously generated on RayStation 11B for all patients to be used in a prior study [23]. Some of them were updated to include improvements on spinal canal and target coverage robustness. All IMPT plans contained 3 beams, placed according to the patient's tumor extension and location. Only one patient needed a range shifter.

PAT plans were created using one arc per plan, and only one revolution around the patient. We chose to use between 1 and 2 degrees as gantry angle spacing, depending on the beam range and amount of total energy layers desired. No range shifter was needed in any case. The TPS allowed us to choose the contralateral lung as a region to be avoided when placing ELs, so this tool was used when possible. Table SM3 contains further planning details.

For both modalities, 4D worst-case minimax robust optimization[24] was performed on a non isotropic CTV expansion, according to Van Herk's formula [25]. Different margins were calculated, depending on the presence or absence of affected lymph nodes.

Although it is well-known that margin recipes fail in proton therapy [26], converting uncertainties into errors in robust optimization is a widely used and pragmatic approach [27]. For consistency with the IMPT plans generated beforehand, and following the work of Badiu et. al [23], the full non-isotropic setup error was divided into two components: 5 mm (isotropic) were included on the RayStation parameter for setup error during robust optimization, while the rest of the setup error was accounted for with a non-isotropic CTV expansion. See Table SM4 for more details. The same clinical goals, as well as trade-off criteria between OAR sparing and target coverage, were applied for both modalities. However, optimization functions were specific to the treatment technique. In both cases, robust objective functions concerned the CTV expansion only. The optimization included a total of 84 scenarios: 7 setup error scenarios (5 mm in 6 different directions, plus nominal case) x 3 range scenarios ($\pm$ 3% range error, plus nominal case) x 4 phases (MidP, EndExh, EndInh, MidV).

Dose distributions in all plans relied on the Monte Carlo dose calculation engine embedded in RayStation. The beam model in use comes from a ProteusPlus machine (IBA s.a.), which allows for 360-degree rotation. The dose calculation grid size was 2.5 x 2.5 x 2.5 $mm^3$. The prescription dose was set to 60 Gy in 30 fractions (2Gy/fx), and all plans were normalized to D50% = 60 Gy.

## 2.4 Plan evaluation and comparison

### 2.4.1 Robustness Evaluation against motion, setup and range error

4D robustness was evaluated on all plans, including MidP, MidV, End Inhale and End Exhale CT scans. Although plan optimization was performed on a CTV expansion (see section 2.3), robustness was evaluated on the raw CTV for preciseness, using a non isotropic setup error equal to the total non isotropic CTV expansion margin used for optimization. Consistently with plan optimization, a 3% range error was used. Considering the four aforementioned phases, 84 robustness evaluation scenarios were analyzed, equaling the number of scenarios used for optimization.

### 2.4.2 Dosimetry analysis

IMPT and PAT plans were compared in terms of predefined clinical goals, as shown in Table 1.

Additionally to relevant CTV and OAR metrics, we computed the homogeneity and conformity index (HI, CI) for the target in the nominal case, as well as body integral dose (ID). The definitions for these metrics can be found in the supplemental material (section SM5).

### 2.4.3 Normal Tissue Complication Probability (NTCP)

NTCPs were calculated as described in the National Indication Protocol of Proton Therapy for Lung Carcinoma from The Netherlands for grade ≥ 2 radiation pneumonitis (NTCP pneumonitis), grade ≥ 2 acute esophageal toxicity (NTCP dysphagia), and 2 years mortality (NTCP 2ym) (see SM6). The detailed NTCP formulas and the clinical parameters involved, such as pulmonary comorbidity, age, or gross tumor volume (GTV) size, can be found in Table SM6. The mean heart dose (MHD), mean lung dose (MLD), and mean esophageal dose (MED) were the dose metrics involved for NTCP 2ym, NTCP pneumonitis, and NTCP dysphagia, respectively. Only the values extracted from the nominal case were used in this calculation.

### 2.4.4 Beam delivery time (BDT) model

For PAT plans, BDT was estimated using ATOM, a BDT simulator algorithm implemented by RaySearch Laboratories [28]. All parameters used for BDT calculation using this model can be found in Table SM2. For IMPT, a simple start-and-stop trajectory was calculated. While human intervention time during IMPT beam delivery was not considered, travel time between beams, including couch and gantry rotation, was accounted for. It should be noticed that, to this date, a clinically approved model to calculate beam delivery time for arc trajectories has not been published.

### 2.5 Interplay simulation

Interplay simulation was performed for all patients and both modalities using OpenTPS, an open source planning software for research purposes [29]. This software relies on MCsquare, an open-source Monte Carlo dose engine for pencil beam scanning [30].

Interplay effect simulation involves a 4D dynamic dose (4DDD) computation process, which requires the 4D-CT phase series, the contours, and the treatment plan (Figure 5). Each simulation starts in a particular phase, and we assume a regular breathing pattern. Details on the process can be found in Figure 1. Using one different starting phase every time, we obtained a total of 10 scenarios per patient, which can then be used to evaluate the interplay effect created by the starting phase variation.

The DVH-band method [4] was chosen to measure plan robustness, by computing bandwidths from perturbed scenarios for CTV D98% and CTV D1% metrics. The narrower the DVH band, the more robust the treatment plan would be (see Figure 6). Beforehand, plans were normalized to D50% = prescription and checked for clinical goal compliance in the nominal case.

## 2.6 Statistical considerations

In the statistical evaluation of setup and range uncertainties, the Wilcoxon signed-rank test in the Python scipy library was used. This test is suitable for a small patient population like ours. Differences were considered non significant if the p-values exceeded 0.05. For interplay effect evaluation, straightforward comparison of bandwidth values was performed, alongside standard deviation calculation for every treatment technique.

# 3. Results

## 3.1 Target coverage and organs at risk

CTV D95% and D98% (Figure 2b) showed no statistical difference (p > 0.05) in the nominal case, meaning that both PAT and IMPT can achieve acceptable target coverage. In fact, both plans complied with the clinical goal for the target. Regarding hot spots in the CTV, revealed by D1%, the techniques did not differ significantly, although the median difference is bigger for PAT plans. In the worst case scenario, CTV D98% showed a wider spread among patients for PAT plans (Figure 2c). However, the median difference with IMPT is negligible (p >> 0.05) and below 1%. In the case of CTV D95%, there is a statistically significant difference (p=0.05) in favor of IMPT, with a median value across all patients equal to 57.9 Gy, against 57.71 Gy for PAT. Nevertheless, the difference is not clinically relevant (< 0.2 Gy), thus allowing PAT plans to preserve target coverage as accurately as IMPT, given the clinical goal for this metric.

As for CTV D1%, our results indicate that high dose spots appear more likely in PAT plans for all the patients, but with a significant statistical difference of 1 Gy only ($p \ll 0.05$). Although the HI index did not show any difference between the two treatment modalities, PAT proved to achieve a higher conformality than IMPT by 66% ($p = 0.00012$).

Regarding the nominal case, PAT plans do not seem to outperform IMPT for any OAR metric median difference, The only significant statistical difference is observed for the heart mean dose (HMD), with PAT plans delivering a higher dose to this organ (23% increased median, $p = 0.00012$). Although spinal canal median D0.04cc in PAT plans is slightly over the one delivered by IMPT, the results are still comparable ($p \gg 0.05$). None of the 28 plans surpasses the clinical goal for this metric. As for the body, PAT plans presented a smaller D1cc ($p = 0.0017$), though there is hardly a 1 Gy difference with IMPT. On the contrary, PAT plans showed an increased body integral dose (see Table SM7a) of 10.62 Gy.cc ($p = 0.00085$) when compared with IMPT.

Looking at the worst-case scenario, the esophagus D0.04cc and HMD exhibit a statistically significant difference in favor of IMPT ($p < 0.05$). However, the analysis of V30% for healthy lung tissue reveals a reduction in radiation dosage for PAT for the majority of patients (Figure 2c), with a median difference of approximately 2 Gy ($p = 0.025$). Although the Spinal Canal does comply with the clinical goal in the worst case scenario for both modalities, the mean of the differences between PAT and IMPT for D 0.04cc presents a non- relevant increase of 3% ($p > 0.05$) in detriment of PAT.

## 3.2 NTCP analysis

In Figure 3, we illustrate the deviation of NTCP values ($\triangle$NTCP) from PAT plans in comparison with IMPT for three types of complications: pneumonitis, dysphagia, and 2-year mortality, which are influenced by the mean doses (in the nominal case) for the lung, esophagus, and heart, respectively (see SM6). As anticipated from the OAR metric results in Figure 2b, NTCP for 2-year mortality increases by 2% (p=0.00012) compared to conventional IMPT. NTCP for pneumonitis and dysphagia of grade 2 also showed approximately a 1% increase with PAT, although without statistical significance.

### 3.3 BDT

The total BDTs for all patients and both modalities are reported in Figure 4, along with their mean values for both modalities (dashed lines). Figure 4 shows that, on average, PAT achieves a slightly shorter BDT, although the difference fails to reach statistical significance (p>0.05) and clinical relevance: PAT mean BDT equals 216.6 seconds versus 227.5 seconds for IMPT. This difference considers the previously defined beam travel time for IMPT as well, which accounts for 55 to 90 seconds maximum (Patient 12). This is not present for arc treatments since there is no lagging time for dynamic arc dose delivery.

### 3.4 Interplay effect evaluation

When analyzing the DVH bandwidths for CTV (Figure 6), D1% bandwidth for PAT plans showed to be slightly larger than for IMPT plans, with a wider data spreading (1.1 $\pm$ 1.4 Gy for PAT vs 1,0 $\pm$ 0.8 for IMPT), hence indicating IMPT plans are more robust when considering high doses within the target. Similarly, PAT showed decreased plan robustness when analyzing target coverage after interplay simulation, with D98% bandwidth = (1.1 $\pm$ 1.3) Gy vs (0.9$\pm$0.9) Gy for IMPT.

# Discussion

Dynamic-ELSA PAT treatment plans yielded clinically acceptable dose distributions, with adequate robustness against setup/range errors and the interplay effect. Compared to IMPT, ELSA-based PAT improved dose conformity but did not outperform the latter for other metrics. Regarding robustness against setup and range errors, target coverage (D98%) was better preserved as anticipated by Seco et al. [31] due to the presence of more irradiation directions. However, if we consider OAR sparing, ELSA PAT plans were less robust than IMPT, as observed in the heart or the esophagus (see Figure 2). A feasible explanation is that these organs were in the path of several beams traversing lung tissue, impacting robustness because of beam passage through density interfaces, all those beams contributing to the total dose distribution. An alternative to improve robustness could be discrete-arc (instead of dynamic) sub plans. This methodology was successfully tested on head and neck cancer [32], showing enhanced robustness. However, further validation is required in thoracic target areas where breathing motion has a higher impact.

A yet unexplored aspect is the use of PTV margins instead of robust optimization, as we assumed that conventional margin recipes may fail due to range uncertainties [26]. This assumption particularly matters when using a few beam directions. However, for wide-angle or full proton arcs, where the range uncertainty is distributed across multiple directions, the effectiveness of PTV-based optimization may not be ruled out. Further research could evaluate the effectiveness of PTV-based versus robust optimization approaches specifically for proton arcs.

The interplay effect was assessed by computing the differences between 10 scenarios per patient and plan. We did not perform a multi fraction dose analysis to observe if this effect was lessened by simulating multiple fractions in a whole treatment. Ideally, we would expect the dose delivered in each fraction to be homogeneous within planning requirements. However, our results showed dose inhomogeneities within a single fraction of the treatment. Consequently, PAT proved to be less robust against interplay, compared to IMPT, not only in terms of target coverage but also regarding hotspot management, with around half of the patient cohort complying with D98% clinical goal for both techniques.

The interplay simulation relies on a regular breathing pattern by means of patient audio-coaching. Implementing breath hold techniques, coupled with visual coaching of the respiratory signal, could improve treatment accuracy. This approach could stabilize the target position, reducing the interplay for PAT or even ruling it out, provided the patient can hold their breath for long enough for the gantry to complete its rotation. While it may be easier to implement in IMPT due to its more straightforward beam delivery, advancements in arc proton therapy could also adopt these techniques, in order to possibly synchronize breathing motion with gantry acceleration and deceleration .

The result obtained here for the HMD should be acknowledged, since it is increased by 23% for PAT in the nominal case (1.4 Gy median difference). On one hand, our result is coherent with the one obtained in a similar previous study [18], where the authors also report a PAT-attributed increased HMD. On the other hand, a compromise between contralateral lung and heart sparing had to be made while planning, given the heart proximity to the target (see Figure 2a). While some planning resources were used to spare it as much as possible, we were limited by hotspot presence. Nevertheless, PAT complies perfectly with the heart clinical dose, not exceeding 20 Gy as mean dose in any case.

Some studies [33][34] have suggested that high total energy deposited in healthy tissues could contribute to the development of secondary cancers and may be a good estimator for quantifying cancer induction. Consequently, the presence of low-dose regions could imply higher risk of secondary malignancies. Although some studies have reported a decrease in integral dose when delivering PAT for target locations such as head and neck [35], as well as breast [17], the results obtained in this article showed a 17% increase when treating lung tumors using ELSA-based PAT. Therefore, this substantial increase may not apply to all target locations and may also depend on the degrees of freedom that the optimiser can explore to reduce low body doses as much as possible.

In a recent review article, Mazal et al. stated that comparing PAT with current standard modalities by looking at dosimetry alone is not enough [36], and that plan evaluation should involve clinical end-points in addition to analyzing dose metrics. Such an assessment plays a pivotal role in informed decision making between the available treatment options [37]. By quantifying and comparing the NTCP differences between the two examined techniques, we could assess the potential of ELSA PAT treatments to mitigate the risk of normal tissue toxicities.

PAT yielded in general worse outcomes for pneumonitis,dysphagia and mortality two years post-treatment for our patient cohort, although statistical significance has been reached for the latter only. The increase in HMD was translated immediately as a 2% (p = 0.00012 ) increment in NTCP-2 year mortality, since the model chosen depends strongly on this metric. A different heart toxicity model could be used for comparison, such as the one mentioned in [38]. According to this article, the 1.4 Gy median difference increase on HMD for our cohort would cause a 10.4% increase in the rate of major coronary events. Taking both results into consideration, and in the context of clinical decision, IMPT would be chosen over PAT for our patient group.

The timing aspect plays a crucial role in clinical proton therapy, with shorter BDT contributing to improved patient comfort, reducing uncertainties in treatment (such as intrafraction motion), and ultimately boosting patient throughput. From Figure 4, we can observe that there are 7 patients (patient 1,2,4,6,8,9 and 13) for which PAT associated BDT is shorter than for IMPT. For all patients except for patient 13, shorter BDT could be related to these plans containing fewer energy layers and spots than their IMPT counterpart (see Table SM8). These patients also presented less complex target location and shape, resulting in less complex plans. Given that ELSA energy layer placement depends solely on geometry, this fact could have helped the optimization algorithm to place energy layers more efficiently. Moreover, the remaining patients with larger BDT for PAT present a higher number of EL and/or spots when compared to IMPT (min 6%, max 47%). Considering the whole cohort, very similar results were obtained in terms of BDT for both modalities, with only a 11 seconds mean advantage of PAT over IMPT. However, if we sum up the time for user interactions, software communications, and security checks performed during IMPT in standard practice, PAT BDT would show a bigger gap. PAT does not require user intervention and performs security checks just once at the start of the arc delivery, hence leading to shorter total treatment time.

Ultimately, a hybrid treatment approach could be considered for lung cancer treatment, combining the best features from IMPT and PAT. IMPT can deliver some treatment fractions, focusing on target control, plan robustness, and decreasing the low dose bath. Meanwhile, other fractions could use PAT to achieve high target dose conformity and effective use of degrees of freedom, potentially improving OAR sparing.

# Conclusion

Although ELSA-based PAT does not outperform conventional IMPT for lung cancer in general, it shows improved target dose conformality while preserving plan quality, although robustness is diminished in the context of non-gated regular breathing assumed in this study. The chosen planning methodology and target proximity to the heart could have contributed to a detriment in the cohort's 2 year mortality due to a substantial increase in the mean dose. The total amount of spots and ELs needed to cover the target efficiently in PAT plans contributes to the small difference in BDT when compared with IMPT, although the lack of beam off and beam switching time for PAT could contribute to enhanced treatment workflow.

## Author contributions

M.C generated and analyzed all the data, wrote the manuscript. S.W is the main developer of OpenTPS, gave advice and provided help with data analysis scripts. D.D helped develop the scripts for interplay simulation in OpenTPS. D.D.P provided the patient database and E.B.V was in charge of curating and cleaning the data. E.E was the main developer of ELSA, provided insight on ELSA-based proton arc planning, reviewed and suggested improvements for the manuscript. J.A.L, A.B.M and E.S supervised the project, corrected and improved the manuscript. E.S gave insight on interplay simulation.

## Acknowledgements

This work was supported by a Walloon Region Mechatech/Biowin grant Nº 8090

## Disclosure statements

Molecular Imaging and Radiation Oncology (MIRO) Laboratory has a research agreement with RaySearch Laboratories.

## Data availability statement

Patient data is not publicly available due to restrictions by our ethical committee, which does not allow data sharing with third parties.

## Ethics declarations

Local ethics committee approval (number 2019/16SEP/402) has been obtained for the collection and use of this retrospective database.

**Tables and figures**

Title: Lung Planning clinical goals

| Structure | Clinical Goal | Worst case |
|---|---|---|
| CTV | D98% >= 57 Gy | D95% >= 57 Gy |
| | D1% < 63 Gy | |
| Oesophagus | D 0.04cc < 60 Gy | - |
| Heart | D mean < 20 Gy | - |
| | D 0.04cc < 60 Gy | - |
| Lungs - GTV | D mean < 20 Gy | - |
| | V30Gy < 20% | - |
| Spinal Canal | D 0.04cc < 50 Gy | D 0.04cc < 50 Gy |
| Body | D1cc < 63 Gy | D1cc < 63 Gy |

Table 1. Clinical goals for IMPT and PAT lung planning. Different criteria were applied for nominal and worst case scenarios.

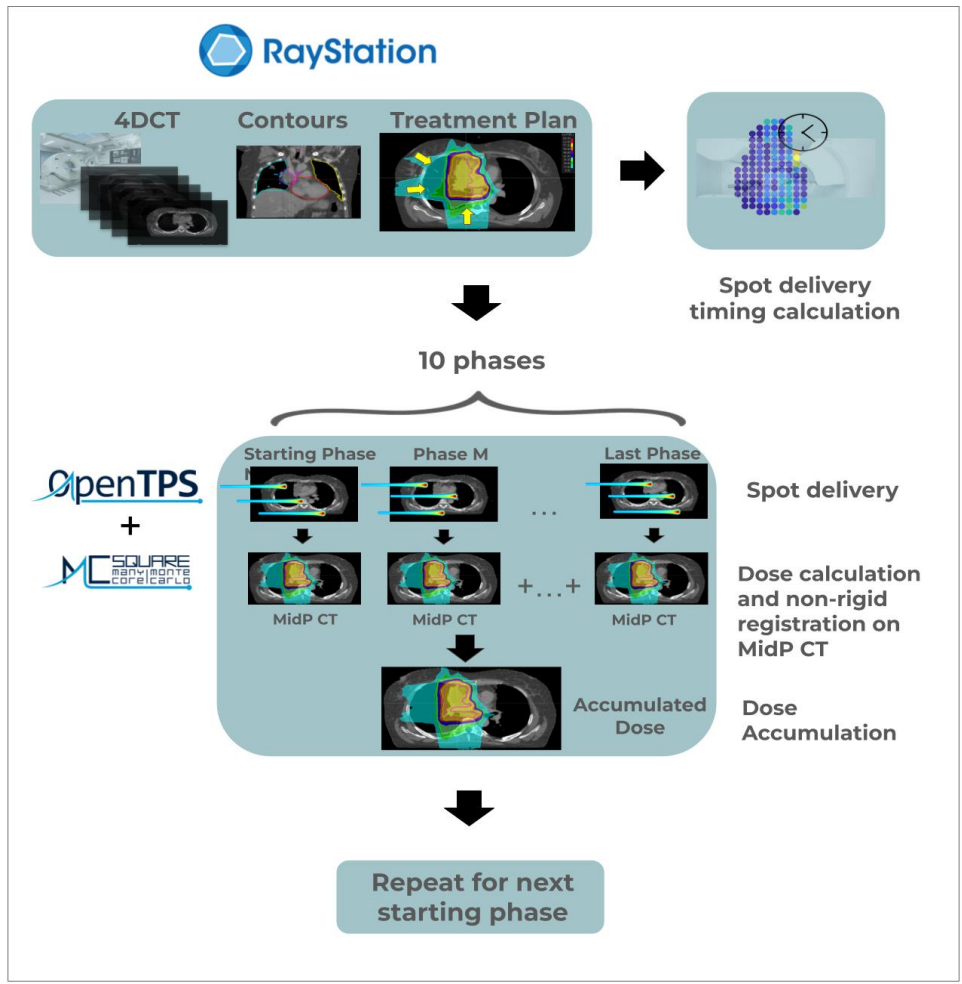

Figure 1. The interplay simulation process in OpenTPS involves utilizing a 4D-CT phase series, contours, and a plan for simulation. Initially, it calculates the timings for delivering proton pencil beam spots given a certain treatment plan and starting phase. Subsequently, the spots in the plan are distributed over the different phases according to the delivery time structure, until the total number of spots have been delivered . The resulting partial dose per phase is computed and then deformed back to the MidPCT image with non-rigid registration. The final outcome involves the summation of all such deformed partial doses, namely, the accumulated dose on MidP. The process is repeated for all phases, where the simulation on each starting phase constitutes one scenario in the interplay simulation, giving a total of 10 scenarios

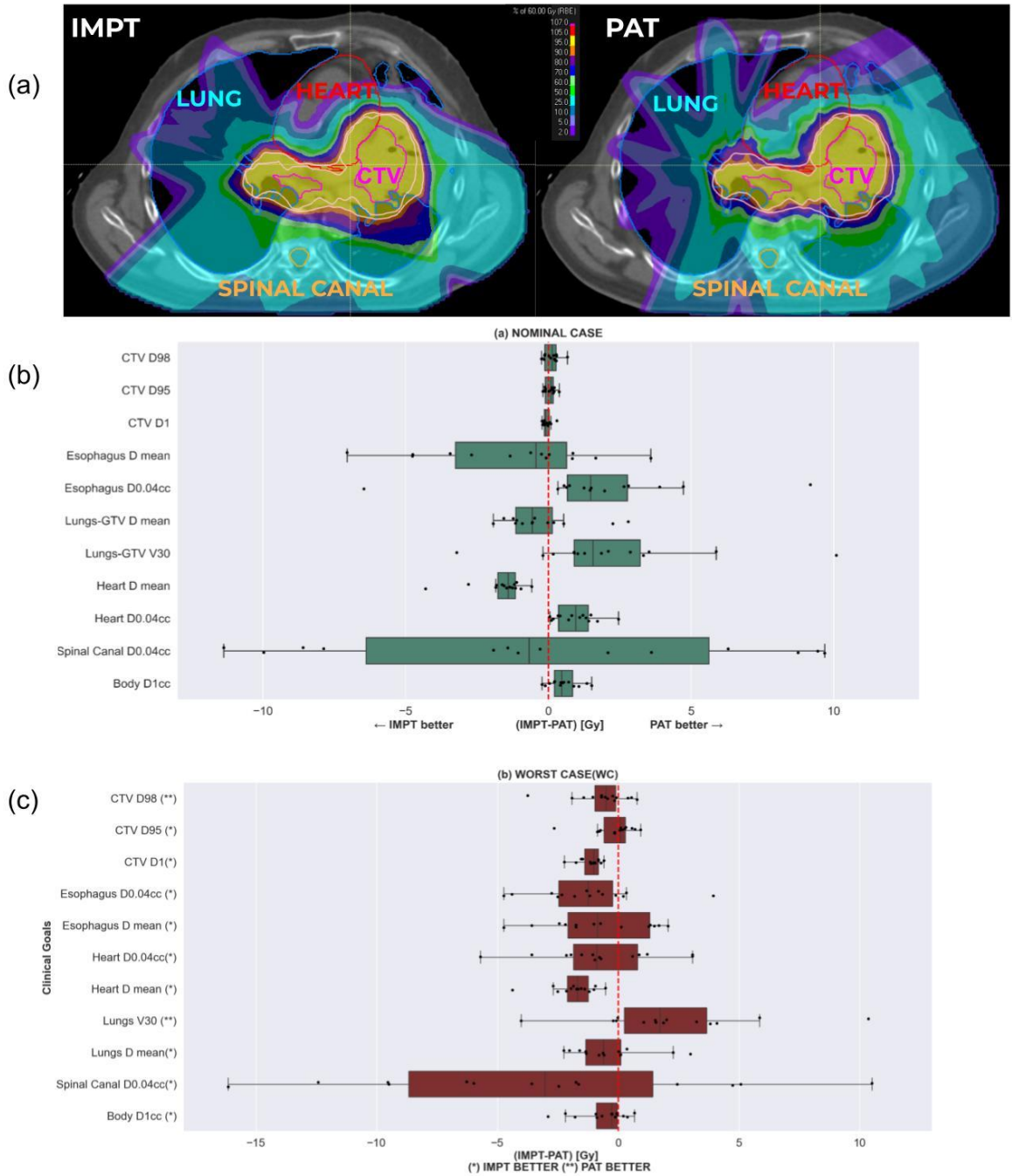

Figure 2. IMPT and PAT plan comparison of setup and range scenarios over all 14 patients. (a) IMPT (left) and PAT (right) Isodose distribution and configuration of OAR with respect to target for one patient (b) Nominal case scenario (c) Worst case scenario. Data points represent fourteen different plan differences, concerning a given metric.

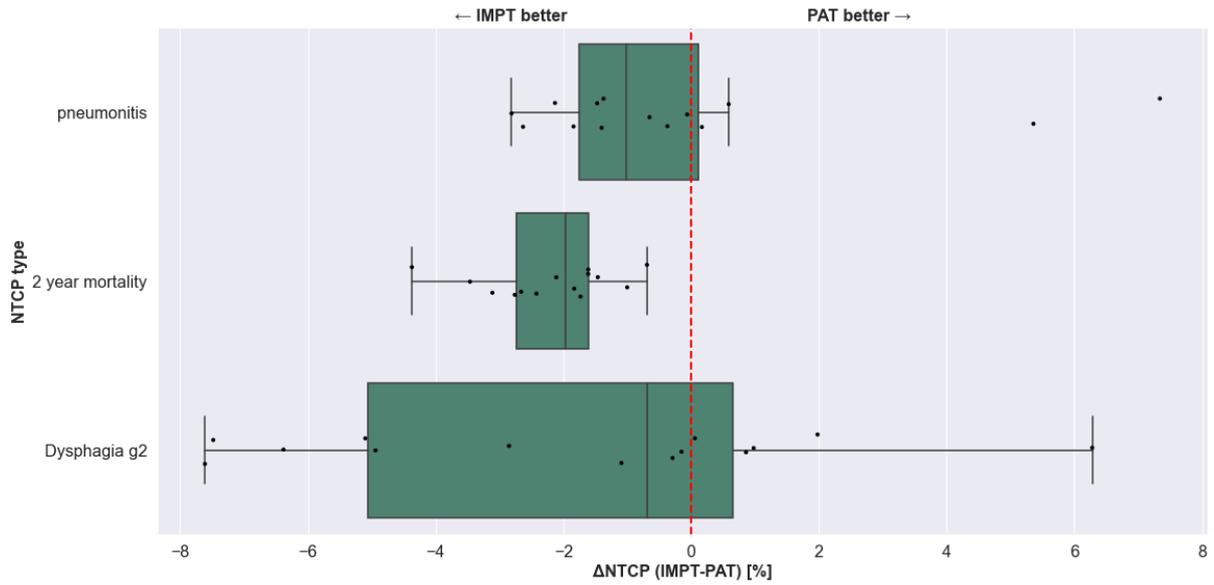

Figure 3. Normal tissue complication probabilities comparison for IMPT and PAT plans. Three types of toxicities are considered: pneumonitis, 2 year mortality and dysphagia grade 2. There is no statistical difference for lung and esophagus toxicities, while 2 year mortality (related to HMD) is increased for PAT plans (p = 0.00012)

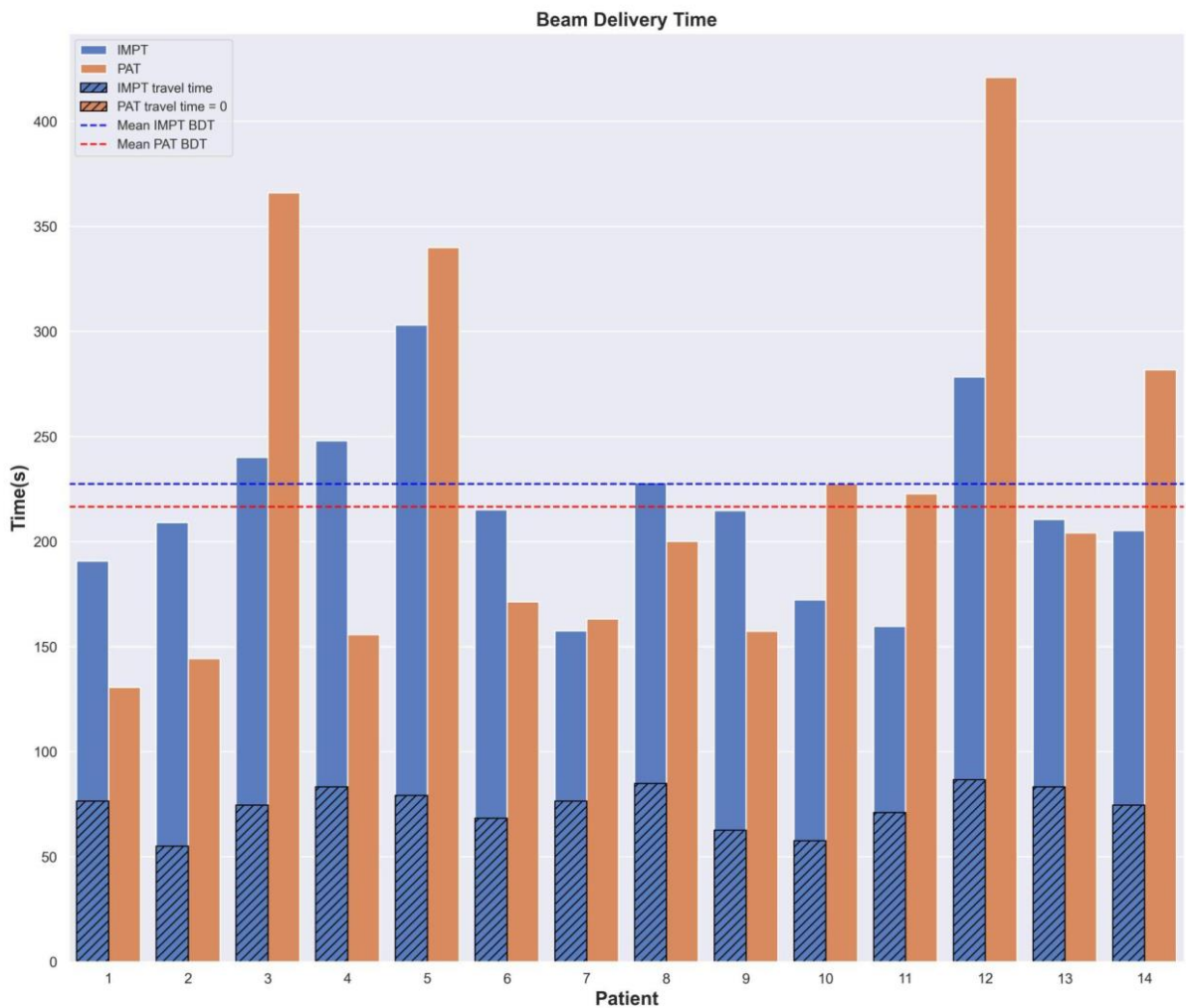

Figure 4. Beam Delivery time comparison between IMPT (blue bars) and PAT (orange bars) plans. Travel time in between beams for IMPT is shown as superimposed dashed bars, while mean IMPT/PAT BDT values are represented by blue and red dashed lines, respectively. Beam delivery time for PAT is mostly similar to IMPT (216.6 s vs 227.5 s, p > 0.05). Notice that PAT travel time does not appear in the plot, since it is always zero.

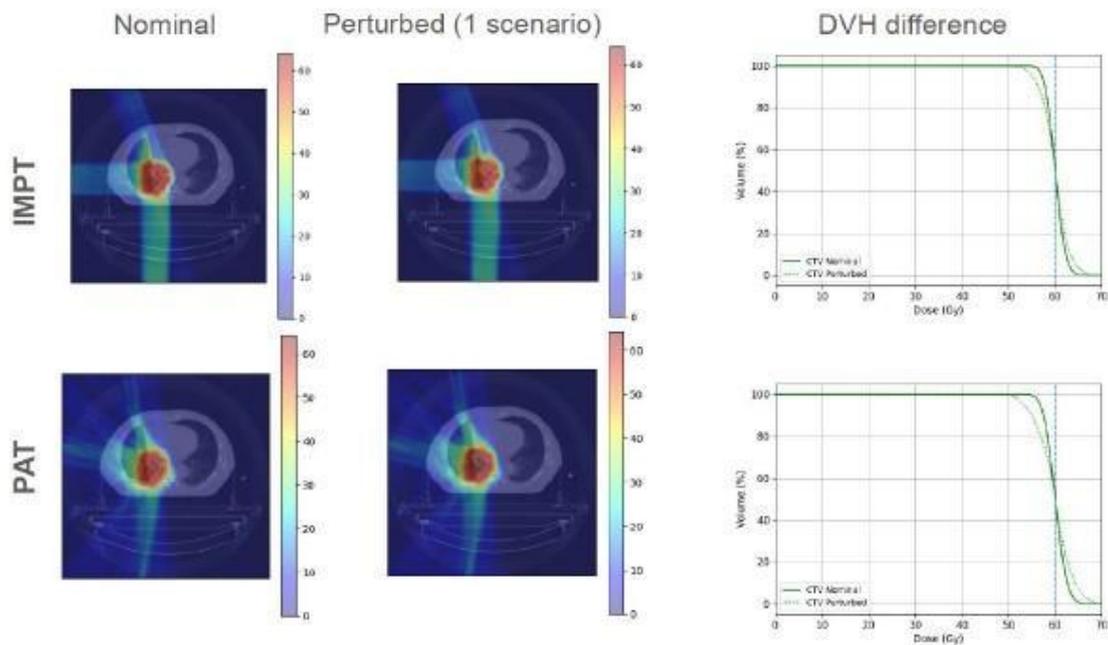

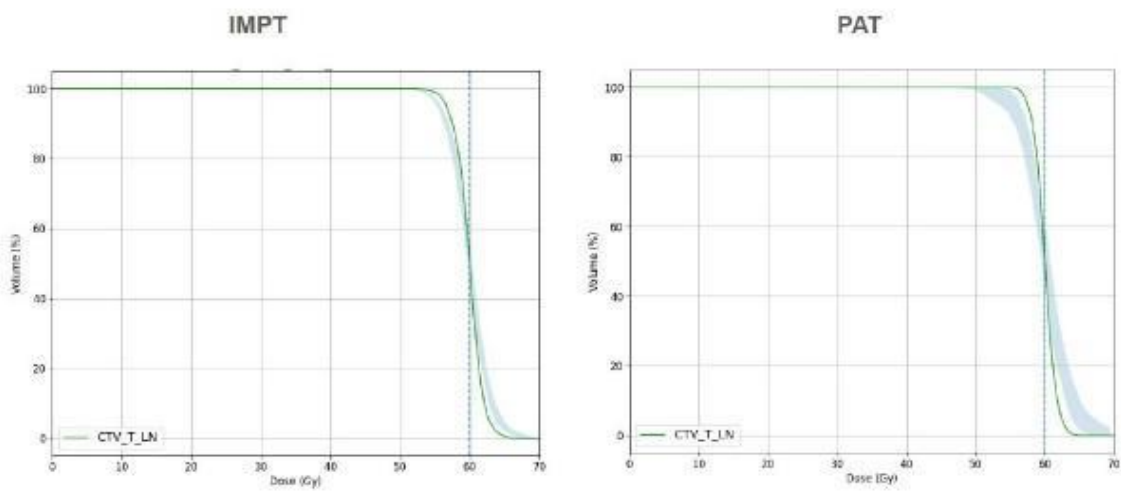

Figure 5.(a) IMPT (upper) and PAT (lower) isodose distribution, together with CTV DVH curve difference, after interplay simulation of one particular scenario for patient 12. (b) CTV DVH bands for patient 12 after interplay simulation of 10 scenarios. Left: IMPT plan, Right: PAT plan. Dashed line indicates the prescription value. All plans normalized to D50% = prescription. For this patient, $\triangle D98\%\ bandwidth\ =\ -3.7\ Gy$, $\triangle D1\%\ bandwidth\ =\ -3.5\ Gy$, $\triangle\ min\ D98\%\ =\ 2.9\ Gy$ $\triangle\ max\ D1\%\ =\ -3.1\ Gy$

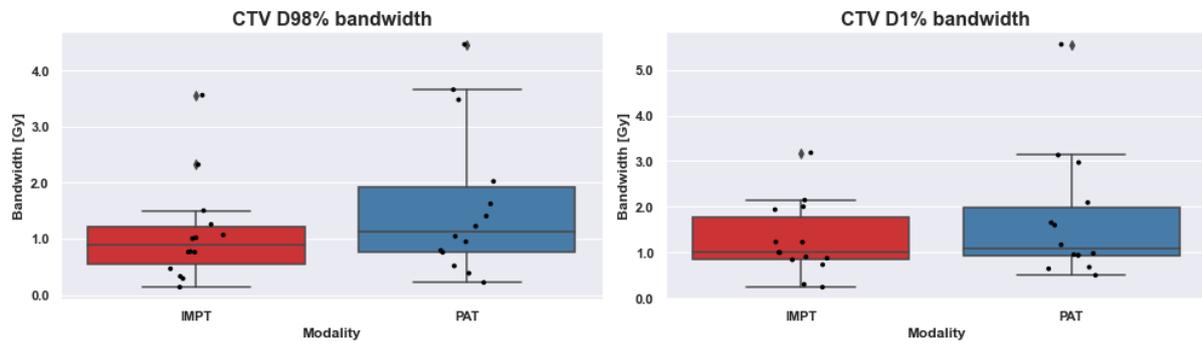

Figure 6. Comparison of interplay simulation on IMPT and PAT plans for 14 patients, over 1 fraction of the treatment and 10 different starting phases. Bandwidths statistics (min - max value) for CTV D98% and CTV D1% are displayed for all fourteen patients, each one represented by a point in the figure

# Supplementary material SM1. Tumor characteristics for the dataset of 14 lung cancer patients.

| Patient | Location | Nodal involvement | CTV tumor size on pCT [cm3] | Tumor motion amplitude [mm] | | |
|---|---|---|---|---|---|---|
| | | | | R-L | A-P | S-I |
| 1 | RML | N | 152.6 | 4.2 | 2.1 | 3.1 |
| 2 | LUL | Y | 165.1 | 1.1 | 1.1 | 2.0 |
| 3 | RML | Y | 331.2 | 0.4 | 0.5 | 1.9 |
| 4 | RUL | Y | 146.4 | 1.2 | 0.7 | 1.6 |
| 5 | RUL | Y | 343.4 | 0.4 | 1.4 | 0.5 |
| 6 | LLL | Y | 177.6 | 3.5 | 3.0 | 3.4 |
| 7 | RUL | Y | 79.1 | 1.3 | 3.5 | 0.7 |
| 8 | RUL | Y | 133.7 | 0.9 | 0.5 | 2.9 |
| 9 | LUL | Y | 442.5 | 0.7 | 0.8 | 0.3 |
| 10 | LUL | Y | 130.2 | 0.8 | 0.9 | 1.5 |
| 11 | RUL | N | 109.6 | 2.2 | 1.8 | 6.6 |
| 12 | RLL | Y | 294.4 | 2.1 | 2.5 | 10.6 |
| 13 | RML | Y | 218.1 | 1.0 | 3.2 | 4.9 |
| 14 | RUL | Y | 268.1 | 4.3 | 2.8 | 3.0 |

Table SM1: Patient characteristics: tumor location, nodal involvement, CTV size and tumor motion amplitude given in absolute value. Abbreviations: LLL = left lower lobe; LUL = left upper lobe; RLL = right lower lobe; RML = right middle lobe; RUL = right upper lobe; Y = Yes; N= No; pCT = Planning CT; R-L = Right-Left; AP = Anterior-Posterior; SI = Superior-Inferior

# Supplementary material SM2. ProteusPlus machine parameters used in the beam delivery time (BDT) calculation model for PAT and IMPT plans.

| | |
|---|---|
| ELST up [s] | 6.0 |
| ELST down [s] | 0.8 |
| Gantry max velocity [deg/s] | 6.0 |
| Gantry max acceleration [s2] | 0.6 |
| Time per spot switch [ms] | 2 |
| Dead time per energy layer [s] | 0.3 |

Table SM2. Relevant machine parameters (ProteusPlus proton therapy machine) for BDT calculation.

# Supplementary Material SM3. IMPT and PAT plan characteristics for the dataset of 14 lung cancer patients

| Patient | IMPT | | | PAT | | |
|---|---|---|---|---|---|---|
| | *Beams (B1,B2,B3) (degrees)* | *Couch kick* | *Range Shifter* | *(arc start, arc stop) (degrees)* | *#Revolutions* | *Gantry angle spacing (degrees)* |
| **1** | (190,240,290) | 0 | N | (310,190) (CCW) | 1 | 2 |
| **2** | (15,90,165) | 0 | Y | (30,170) (CW) | 1 | 2 |
| **3** | (220,270,310) | 0 | N | (180,0) (CW) | 1 | 1 |
| **4** | (200,270,340) | 0 | N | (330,190) (CCW) | 1 | 2 |
| **5** | (325,270,210) | 0 | N | (180,0) (CW) | 1 | 1 |
| **6** | (110,170,230) | 0 | N | (50,210) (CW) | 1 | 2 |
| **7** | (190,240,300) | 0 | N | (340,200) (CCW) | 1 | 2 |
| **8** | (190,240,340) | 0 | N | (0,180) (CCW) | 1 | 2 |
| **9** | (120,160,200) | 0 | N | (20,160) (CW) | 1 | 2 |
| **10** | (90,130,170) | 0 | N | (30,170) (CW) | 1 | 1,5 |
| **11** | (170,210,250) | 0 | N | (320,190) (CCW) | 1 | 1 |
| **12** | | 0 | N | (180,0) (CW) | 1 | 1 |
| **13** | (170,270,220) | 0 | N | (190,300) (CW) | 1 | 1 |
| **14** | (180,225,270) | 0 | N | (180,0) (CW) | 1 | 1 |

Table SM3. Plan characteristics: beam angles (IMPT), couch rotation, presence of a range shifter, arc range (PAT), number of revolutions around the patient and gantry angle spacing. Abbreviations: B1,B2,B3 = beam 1, 2 and 3, CW = clockwise, CCW = counterclockwise.

# Supplementary material SM4. CTV Margin calculation for robust optimization

Maximum setup errors were calculated using systematic ($\Sigma$) and random ($\sigma$) setup and baseline shift values were used for two cases: tumor only and tumor with lymph nodes. This was done using van Herk's margin formula (1) with the goal of obtaining a margin that ensures a minimum dose is delivered to 90% of the patient population:

$$M_{PTV} = 2.5\Sigma_{total} + 0.7\sigma_{total} \quad (1)$$

|  | $X_{(sagittal)}$ (mm) | $Y_{(coronal)}$ (mm) | $Z_{(transverse)}$ (mm) |
|---|---|---|---|
| $\Sigma_{BL}$ | 1.8 | 1.6 | 1.9 |
| $\Sigma_S$ | 1.6 | 2 | 2.4 |
| $\sigma_{BL}$ | 1.6 | 1.6 | 2.1 |
| $\sigma_S$ | 1.8 | 2.1 | 2.1 |
| $\Sigma_{total}$ | 2.41 | 2.56 | 3.06 |
| $\sigma_{total}$ | 2.41 | 2.64 | 2.97 |
| $M_{PTV}$ | 7.71 | 8.25 | 9.73 |
| Expanded CTV margin | 2.71 | 3.25 | 4.73 |

Table SM4.1 Setup errors for the case with only primary tumor

|  | $X_{(sagittal)}$ (mm) | $Y_{(coronal)}$ (mm) | $Z_{(transverse)}$ (mm) |
|---|---|---|---|
| $\Sigma_{BL}$ | 1.9 | 1.6 | 1.9 |
| $\Sigma_S$ | 1.6 | 2 | 2.4 |
| $\sigma_{BL}$ | 1.7 | 1.6 | 2.1 |
| $\sigma_S$ | 1.8 | 2.1 | 2.1 |
| $\Sigma_{total}$ | 2.48 | 2.56 | 3.06 |
| $\sigma_{total}$ | 2.48 | 2.64 | 2.97 |
| $M_{PTV}$ | 7.94 | 8.25 | 9.73 |
| Expanded CTV margin | 2.79 | 3.25 | 4.73 |

Table SM4.2 Setup errors for the case with primary tumor and lymph nodes

The calculated total setup error values for each direction were inputted in our treatment planning system (TPS), RayStation. However, using a margin that exceeds 5 mm in RayStation leads to a significant increase in optimization time due to the system automatically generating intermediate errors to ensure robust coverage, which can be an over-conservative approach. As a workaround, instead of optimizing on the CTV using the full setup error, a patient-specific CTV expansion can be generated to account for part of the setup error. From the full value of the margin, 5 mm is subtracted (in each direction) and subsequently used as an isotropic setup error value. The CTV is then expanded by the remaining amount. For instance, for a setup error of 7.71 mm in the sagittal direction, a CTV expansion of 2.71 mm was robustly optimized with a setup error of 5 mm (7.71 mm = 2.71 mm in the margin + 5 mm in the robust optimization).
The complete setup margin values can be found in the Appendix (Table SM3.1 and Table SM3.2). Optimization was done on the expanded CTV volume to reduce computation time.

The total number of optimization scenarios was 84: 7 (setup errors: ± 5 mm in x,y,z directions, additionally the nominal scenario) × 3 (image conversion errors: ±3%, 0%) × 4 (breathing phases: MidP, maximum inhale, maximum exhale and an additional mid-ventilation phase).

## Supplementary Material SM5. Target homogeneity index (HI), conformity index (CI) and body integral dose (ID) definitions for IMPT and PAT plan dosimetry assessment

Target dose homogeneity was evaluated with the Homogeneity index (HI), defined as a slight variation of the formula in [1]

$$HI = \frac{(D1\% - D98\%)}{D\ prescribed}.$$

Target conformity was assessed as well, by means of the Conformity index (CI), defined as [2]

$$CI = \frac{V95\%}{V_T},$$

where $V95\%$ is the volume covered by 95% of the prescribed dose, and $V_T$ is the total volume of the target. The body integral dose was assessed as well, defined as

$$ID = \underline{D}.V,$$

where $\underline{D}$ is the mean dose to the body in Gy, while $V$ is the patient volume in cc.

## Supplementary material SM6. Dutch Normal Tissue Complication Probability (NTCP) model and patient characteristics for the dataset of 14 lung cancer patients.

$$NTCP_x = 1/(1 + e^{-S(x)})$$

**S(Grade ≥2 pneumonitis)** = −4.12 + 0.138 ∗ MeanLungDose − 0.3711 ∗ (Smoking: stopped) − 0.478 ∗ (Smoking: active) + 0.8198 ∗ (Pulmonary comorbidity) + 0.6259 ∗ (Tumor location) +
0.5068 ∗ Age + 0.47 ∗ (Sequential chemotherapy)

**S(Grade ≥2 acute esophageal toxicity)** = -3.634 + 1.496*ln(MeanEsophagealDose) - 0.0297*(Interval start-stop RT)

**S(2 year mortality)** = −1.3409 + 0.0590 ∗ $SQRT$(GTVvolume) +0.2635 ∗ SQRT(MeanHeartDose)

| Patient | Dysphagia Gr2 Overall treatment time (days) | Pneumonitis Smoking Stopped Smoking * | Active smoking ** | Pulmonary Comorbidity*** | Lobe**** | Seq. Chemotherapy ***** | Age |
|---|---|---|---|---|---|---|---|
| 1 | 33 | 1 | 0 | 0 | 1 | 0 | 70 |
| 2 | 41 | 1 | 0 | 1 | 0 | 0 | 79 |
| 3 | 40 | 1 | 0 | 1 | 1 | 0 | 64 |
| 4 | 40 | 0 | 1 | 0 | 0 | 0 | 57 |
| 5 | 44 | 0 | 1 | 1 | 0 | 0 | 62 |
| 6 | 39 | 1 | 0 | 1 | 1 | 0 | 74 |
| 7 | 42 | 0 | 0 | 0 | 0 | 0 | 71 |
| 8 | 40 | 1 | 0 | 0 | 0 | 0 | 79 |
| 9 | 40 | 0 | 1 | 1 | 0 | 0 | 58 |
| 10 | 40 | 1 | 0 | 0 | 0 | 0 | 78 |
| 11 | 40 | 1 | 0 | 0 | 0 | 0 | 70 |
| 12 | 40 | 0 | 1 | 1 | 1 | 0 | 71 |
| 13 | 40 | 1 | 0 | 1 | 1 | 0 | 73 |
| 14 | 40 | 1 | 0 | 1 | 0 | 0 | 58 |

Table SM6. Clinical characteristics of patients considered for the NTCP calculation. * stopped smoking = 1, never or active smoker = 0 ; ** active smoker = 1, never smoked or quit = 0 ; *** COPD or other pre-existing lung disease = 1, None = 0 ; **** Middle/lower lobe = 1, upper lobe = 0 ; ***** Yes = 1 , No = 0.

For more details, see [3].

## Supplementary material SM7. Dosimetric and NTCP results on plan comparison between PAT and IMPT, for the dataset of 14 lung cancer patients

| Clinical Goal | IMPT Median (NC)[Gy] | Median (WC)[Gy] | PAT Median (NC)[Gy] | Median (WC)[Gy] | IMPT-PAT Median (NC)[Gy] | p-value | Median (WC)[Gy] | p-value |
|---|---|---|---|---|---|---|---|---|
| CTV D95% | 58,99 | 57,90 | 58,90 | 57,71 | 0,09 | 0,15 | 0,07 | 0,86 |
| CTV D98% | 58,77 | 56,48 | 58,61 | 57,00 | 0,14 | 0,12 | -0,49 | 0,05 |
| CTV D1% | 61,37 | 62,09 | 61,50 | 63,05 | -0,09 | 0,14 | -0,99 | 0,00012 |
| CTV HI | 0,04 | - | 0,05 | - | 0,00 | 0,19 | - | - |
| CTV CI | 4,36 | - | 2,62 | - | | 0,00012 | - | - |
| Esophagus D mean | 22,79 | 27,83 | 23,64 | 28,04 | -0,43 | 0,19 | -0,86 | 0,27 |
| Esophagus D0.04 cc | - | 63,74 | - | 65,50 | 1,43 | 0,0100 | -1,75 | 0,02 |
| Heart D mean | 5,99 | 9,81 | 7,29 | 10,84 | -1,41 | 0,00012 | -1,70 | 0,00012 |
| Heart D0.04 cc | - | 64,97 | - | 65,50 | 1,05 | 0,00012 | -0,58 | 0,33 |
| Lungs - GTV D mean | 14,12 | 14,85 | 13,10 | 15,07 | -0,57 | 0,24 | -0,62 | 0,36 |
| Lungs V30Gy | 17,46 | 20,37 | 16,04 | 18,79 | 1,56 | 0,0085 | 1,69 | 0,025 |
| Spinal Canal D0.04cc | 28,97 | 36,68 | 31,80 | 47,57 | -0,68 | 0,952 | -3,13 | 0,14 |
| Body D1cc | 62,61 | 64,16 | 62,09 | 64,77 | 0,50 | 0,0017 | -0,23 | 0,15 |
| Body ID [Gy.cc] | 121,80 | - | 147,17 | - | -10,62 | 0,00085 | - | - |

Table SM7a. Median values per treatment technique, and median values for the difference. NC = nominal case, WC = worst case, ID = integral dose

| Strategy | NTCP_pneumonitis | NTCP_2y_mortality | NTCP_dysphagia_g2 |
|---|---|---|---|
| IMPT | 11.3 | 45.61 | 44.79 |
| arcPT | 13.185 | 47.745 | 47.185 |

Table SM7b. Median NTCP values per treatment technique.

| NTCP metric | Median Delta NTCP (IMPT-PAT) | p value |
|---|---|---|
| NTCP_2y_mortality | -1.97 | 0,00012 |
| NTCP_dysphagia_g2 | -0.69 | 0,14 |
| NTCP_pneumonitis | -1.01 | 0,24 |

Table SM7c. Median values for the difference in NTCP between the treatment modalities

## Supplementary Material SM8. IMPT and PAT plan characteristics influencing BDT results, and delivery time values for the dataset of 14 lung cancer patients

| | Number of beams | | total EL | | ELS up | | ELS down | | #Spots | | BDT(s) | |
|---|---|---|---|---|---|---|---|---|---|---|---|---|
| Patient | IMPT | PAT | IMPT | PAT | IMPT | PAT | IMPT | PAT | IMPT | PAT | IMPT | PAT |

| | | | | | | | | | | | |
|---|---|---|---|---|---|---|---|---|---|---|---|
| 1* | 3 | 1 | 97 | 61 | 2 | 5 | 94 | 55 | 14026 | 11933 | 190.8 | 130.7 |
| 2* | 3 | 1 | 120 | 71 | 2 | 5 | 117 | 65 | 22633 | 10438 | 209.1 | 144.3 |
| 3 | 3 | 1 | 110 | 180 | 2 | 17 | 107 | 163 | 27826 | 28745 | 240.1 | 366.1 |
| 4* | 3 | 1 | 112 | 71 | 2 | 5 | 109 | 65 | 26955 | 15817 | 247.9 | 155.7 |
| 5 | 3 | 1 | 132 | 181 | 2 | 7 | 129 | 173 | 43684 | 43997 | 303.1 | 339.9 |
| 6* | 3 | 1 | 110 | 81 | 2 | 5 | 107 | 75 | 20469 | 18622 | 215.2 | 171.4 |
| 7 | 3 | 1 | 82 | 71 | 2 | 9 | 79 | 61 | 7653 | 9142 | 157.5 | 163.1 |
| 8* | 3 | 1 | 110 | 91 | 2 | 7 | 107 | 83 | 20123 | 19811 | 228.2 | 200.1 |
| 9* | 3 | 1 | 100 | 71 | 2 | 4 | 97 | 66 | 25241 | 19699 | 214.8 | 157.3 |
| 10 | 3 | 1 | 88 | 94 | 2 | 11 | 85 | 82 | 15527 | 23132 | 172.3 | 227.6 |
| 11 | 3 | 1 | 77 | 131 | 2 | 6 | 74 | 124 | 10276 | 17389 | 159.6 | 222.7 |
| 12 | 3 | 1 | 117 | 181 | 2 | 24 | 114 | 156 | 33561 | 32451 | 278.3 | 420.9 |
| 13* | 3 | 1 | 98 | 111 | 2 | 5 | 95 | 105 | 18028 | 21472 | 210.6 | 204.1 |
| 14 | 3 | 1 | 96 | 181 | 2 | 5 | 93 | 175 | 18912 | 20969 | 205.2 | 281.8 |

Table SM8. Plans characteristics: number of beams or arcs, total number of energy layers, energy layer switchings, number of spots and beam delivery time results. Abbreviations: IMPT = intensity modulated proton therapy, PAT = proton arc therapy, EL = energy layers, ELSU = energy layer switching up, ELSD = energy layer switching down, BDT = beam delivery time. * indicate that the PAT plan held a smaller BDT value than the IMPT plan.